# Peak current reduction in presence of RF phase modulation in the dual RF system

Bocheng Jiang[1], Yao Zhang[1*], Cheng-Ying Tsai[2]

1. Laboratory for Ultrafast Transient Facility, Chongqing University, Chongqing 401331, China.
2. School of Electrical and Electronic Engineering, Huazhong University of Science and Technology, Wuhan 430074, China

2025/2/12

**Abstract**

High harmonic cavities are widely used in electron storage rings to lengthen the bunch, lower the bunch peak current, thereby enhancing the Touschek lifetime, reducing the IBS effect, as well as providing Landau damping. When the RF phase is modulated in such a dual-RF system, simulations reveal that the peak current can be further reduced under a wide modulation frequency range. This paper presents the simulation results and explains the underlying mechanism.

## I.   Introduction

Low-energy electron storage rings suffer from the Touschek effect and intra-beam scattering (IBS), which can lead to a significantly shortened beam lifetime and substantial emittance growth [1-7]. This issue persists even in medium-energy storage rings when the emittance becomes extremely small, as seen in diffraction-limited light sources [8-10].

A common strategy to mitigate these effects is to reduce the charge density within a bunch. This can be achieved through two primary methods: increasing transverse coupling to expand the vertical beam size or lengthening the bunch. For the latter, high-harmonic cavities are widely implemented in storage rings to flatten the accelerating voltage gradient at the synchrotron phase, thereby increasing the bunch length [11-15]. These cavities can operate in either an active or passive mode.

Another effective approach is RF modulation, which can be applied to either the phase or voltage at the twice or triple times of the synchrotron frequency. Such modulation drives particle oscillations to larger amplitudes, thereby stretching the bunch length. The use of RF modulation for bunch lengthening has been extensively studied both analytically and experimentally [16-20], most investigations have focused on single-RF systems. To our knowledge, only one study has explored phase modulation on a dual-RF system — conducted at the Indiana University Cyclotron Facility (IUCF) proton Cooler Ring [21,22]. In that study, the diffusion mechanism induced by phase modulation was revealed, and the resulting bunch lengthening effect was observed.

Compared to a proton storage ring, the beam parameters in an electron storage ring are determined by quantum excitation and radiation damping. The impact of RF modulation on a dual-RF system is particularly intriguing—if it can further extend the bunch length, it would significantly enhance the performance of storage rings, especially for the 4[th] generation light sources. In this paper, we investigate this issue through simulations and provide a preliminary analysis.

## II.   Simulation of RF modulation

A tracking code has been developed to simulate the longitudinal beam dynamics. The turn-by-turn evolution of the longitudinal motion is described by Eqs. (1–4). For simplicity, only the first-order momentum compaction is considered in the tracking process. The storage ring beam parameters used in the simulation are listed in Table 1 [23]. The first-order momentum compaction factor for this ring is 8.15×10⁻³, which is relatively large. As a result, the tracking remains accurate without the need to include higher-order momentum compaction terms, since the first-order effect dominates. In the meantime, to maintain a 3% linear momentum acceptance, the RF voltage should be 750 kV which results to an original short bunch length of 2.4mm.

* zhang_yao@cqu.edu.cn

$$\varphi_{rf}^{n+1} = \varphi_s + \omega_{rf} dt^n + A_{rf} Sin(\omega_{mod} n T_0) \quad (1)$$

$$\varphi_{rf3}^{n+1} = \varphi_3 + \omega_{rf3} dt^n + A_{rf3} Sin(\omega_{mod3} n T_0) \quad (2)$$

$$dE^{n+1} = dE^n - U_0 + V_{rf} Cos(\varphi_{rf}^{n+1}) + V_{rf3} Cos(\varphi_{rf3}^{n+1}) - \frac{2T_0}{\tau_s} dE^n + rand(1)\varepsilon_c\sqrt{\frac{11N_p}{27}} \quad (3)$$

$$dt^{n+1} = dt^n + \frac{\alpha T_0}{E_0} dE^{n+1} \quad (4)$$

Where $\varphi_{rf}$, $\varphi_{rf3}$ are the phases of the main and the 3rd harmonic cavity respectively (here subscript 3 denotes the 3rd harmonic cavity and henceforth), $\varphi_s$ is the synchrotron phase, $\varphi_3$ is the initial phase of 3rd harmonic cavity, $\omega_{rf}$, $\omega_{rf3}$ are the RF angular frequencies, $A_{rf}$, $A_{rf3}$ are the amplitudes of the phase modulation, $\omega_{mod}$, $\omega_{mod3}$ are the angular frequencies of the phase modulation. $dE$ is the energy deviation, $U_0$ is the energy loss per turn, $V_{rf}$, $V_{rf3}$ are the peak RF voltages, $T_0$ is the revolution period, $\tau_s$ is the longitudinal damping time, $rand(1)$ is a random number with standard deviation equal to 1, $\varepsilon_c$ is the critical photon energy of the bend radiation, $N_p$ is the photon numbers emitted per turn, $\alpha$ is the momentum compaction, $E_0$ is the beam energy.

The 5th term in the right hand of Eq.(3) is the radiation damping, and the 6th term is the quantum excitation. Where the critical energy of the radiation and the photon numbers per turn can be calculated by the following equation[24]:

$$\varepsilon_c(\text{keV}) = 0.665[E_0(GeV)]^2 B(Tesla) \quad (5)$$

$$N_p = \frac{15\sqrt{3}}{8} \frac{U_0}{\varepsilon_c} \quad (6)$$

Table 1. Beam parameters for simulation

| Parameters | Value |
| --- | --- |
| Beam energy (GeV) | 0.5 |
| Ring circumference (m) | 76.78 |
| Momentum compaction | 8.15e-3 |
| Energy loss per turn (keV) | 4.34 |
| Initial Energy spread | 0.038% |
| Initial bunch length (mm) | 2.4 |
| Main RF frequency (MHz) | 499.8 |
| Main RF voltage (kV) | 750 |
| Longitudinal damping time (ms) | 29.5 |
| Bending magnet field (T) | 1.31 |

In the simulation, 1×10⁴ macro-particles are used and 5×10⁴ turns are tracked. When only the main RF system is applied, the RMS bunch length and energy spread agree well with the theoretical values when radiation damping and quantum excitation reach equilibrium, as shown in Fig. 1. This study primarily focuses on the peak current (longitudinal charge density), therefore, we count macro-particles in the longitudinal coordinate as a representation of the peak current, as illustrated in Fig. 1(c).

* zhang_yao@cqu.edu.cn

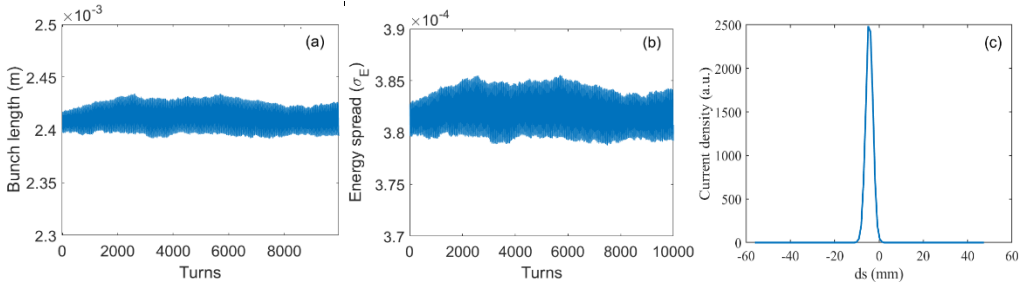

Fig. 1. RMS bunch length (a), energy spread (b) and particle distribution in longitudinal coordinate (c).

When a 3$^{rd}$ harmonic cavity is introduced with a voltage of 250 kV, the bunch is stretched, resulting in a flattened beam current profile with a peak current approximately 1/7 of single RF cavity case, as shown in Fig. 2(a). The corresponding phase-space distribution and synchrotron tune spread are depicted in Fig. 2(b, c).

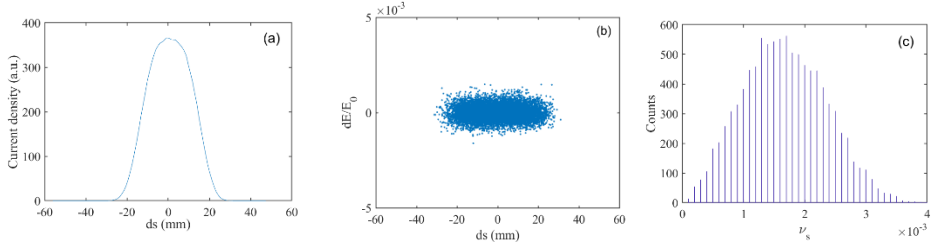

Fig. 2. 3$^{rd}$ harmonic cavity working in an optimized status, (a) particle longitudinal distribution, (b) longitudinal phase space, (c) synchrotron tune spread.

When the 3$^{rd}$ harmonic cavity voltage is increased to 268 kV, the bunch becomes over-stretched, forming a double-Gaussian distribution, as shown in Fig. 3(a). In this case, the RMS bunch length alone cannot fully characterize the beam performance, because effects such as Touschek lifetime and IBS are more sensitive to charge density. This is why we use peak current as the indicator. The corresponding phase-space distribution and synchrotron tune spread are presented in Fig. 3(b) and (c), respectively.

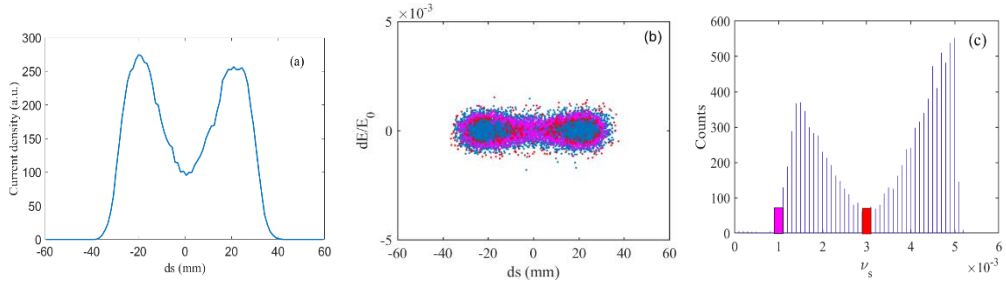

Fig. 3. 3$^{rd}$ harmonic cavity working in an overstretched status (The electrons plotted in magenta and red dot in (b) has the frequency indicated by the color bar in (c) ).

* zhang_yao@cqu.edu.cn

When $V_{rf3}$=268kV, $\omega_{mod3} = 2\pi \times 30.46$ kHz ($\omega_{mod3}/\omega_0 = 0.008$) and $A_{rf3} = 2°$, the bunch length is further increased, and the peak current is reduced to approximately half of that in Fig. 3(a), as shown in Fig. 4(a). The longitudinal phase space occupied by the particles is depicted in Fig. 4(b). It is observed that not only does the bunch length increase, but the energy spread also more than doubles, reaching 0.092%.

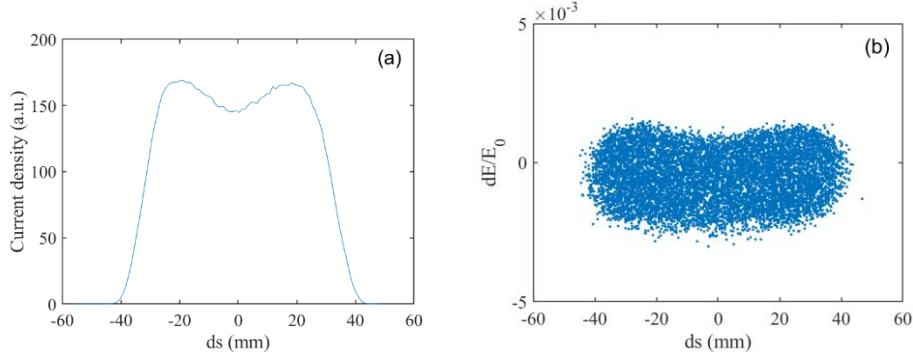

Fig. 4. Beam current profile and phase space for the case of dual RF over stretched case with RF phase modulation.

For a single RF system, the modulation frequency should be close to $2\omega_s$ or $3\omega_s$. However, in a dual-RF system, the synchrotron tune is spread. It is necessary to find out whether effective modulation can occur over a broader frequency range. The results, as shown in Fig. 5, confirm this expectation—modulation remains effective in reducing peak current across a wide frequency band. Since the synchrotron tune is spread, it is hard to say the modulation frequency is $2\omega_s$ or $3\omega_s$. For dual RF system, the RF modulation is more feasible as the modulation frequency does not need to be precisely set. Also, for a broad frequency range, the modulation won't cause beam loss.

The error bar in Fig. 5 comes from different peak current at different turns, which means the bunch is oscillating in longitudinal phase space, the oscillations can be suppressed at proper modulation frequency as shown in Fig. 5.

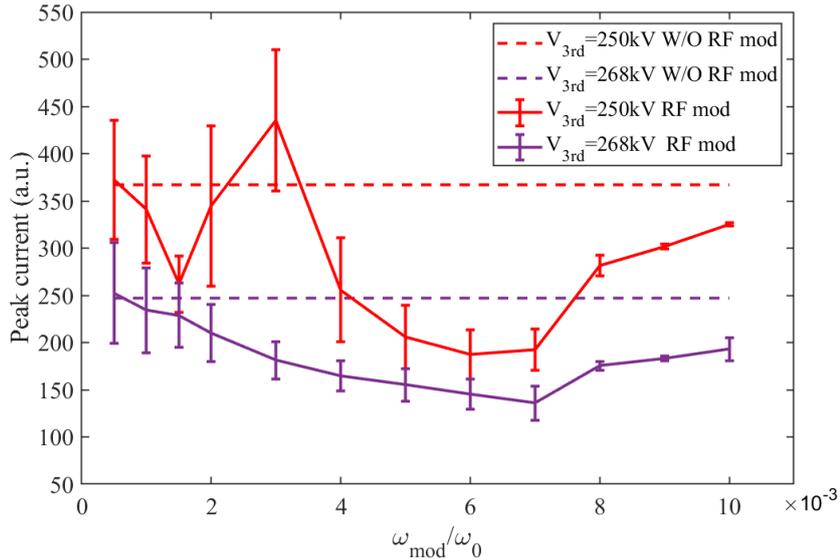

Fig. 5. RF modulation at different frequencies.

### III. Analysis of RF modulation

The theoretical explanation for dual-RF system with RF modulation can follow Ref. [22], it predicted a large number of resonance islands around stable fixed point exist. The island width is approximately given by:

$$\Delta I \approx 4 \left[\frac{v_m a |g_n|}{\left|\frac{\partial Q_s}{\partial I}\right|}\right]^{1/2} \quad (7)$$

* zhang_yao@cqu.edu.cn

Here $v_m$ and $a$ are the modulation tune and the modulation amplitude respectively, $Q_s$ is the synchrotron tune, $g_n(J) = \frac{1}{2\pi}\int_{-\pi}^{\pi} \delta e^{-in\psi} d\psi$ is the expansion amplitude.

Compared to the RF modulation studied in IUCF proton Cooler Ring, the longitudinal phase space occupied by the electron beam is much smaller than that of the proton beam, and the modulation frequency is much higher. Making a larger island width in a small phase space, the island is more likely to overlap with each other. Bunch lengthening will occur due to overlapping parametric resonances allowing electrons diffuse into the many surrounding resonance islands.

The trajectories in phase space of two electrons with different initial coordinates are plotted in Fig. 6, for this plot the radiation damping and quantum excitation are turned off in Eq. (3), and the turns for tracking are extended to $2\times10^5$. It is found that the electron will not stay in an island for a long time then drift away, the trajectories seems in chaos occupy in a larger area.

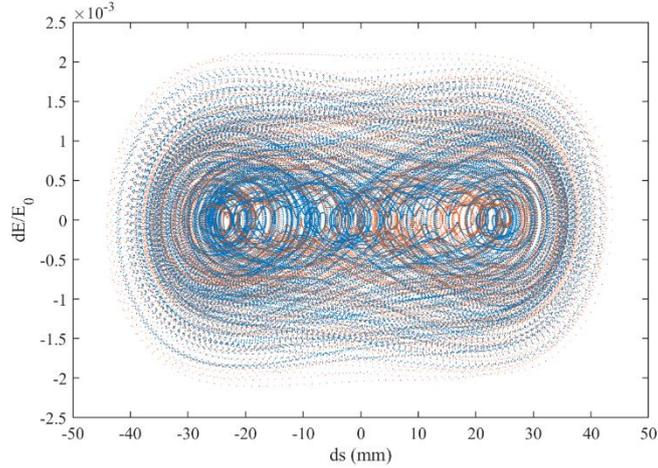

Fig. 6. Electron trajectories in phase space without radiation damping and quantum excitation.

## IV. Summary

The simulation reveals that peak current can be further reduced for dual-RF system in presence of phase modulation, which will benefit the 4[th] generation light sources as long as the increased energy spread is in an acceptable range. The modulation frequency does not need to be precisely set which makes the method more feasible in real machines. Further investigation is required to understand the impact of chaotic motion in the longitudinal plane, along with the reduced peak current and increased energy spread, on beam instabilities.

* zhang_yao@cqu.edu.cn

* zhang_yao@cqu.edu.cn

* zhang_yao@cqu.edu.cn